# Congestion Control Protocol for Wireless Sensor Networks Handling Prioritized Heterogeneous Traffic

Muhammad Mostafa Monowar, Md. Obaidur Rahman, Al-Sakib Khan Pathan, and
Choong Seon Hong
Department of Computer Engineering, Kyung Hee University
1 Seocheon, Giheung, Yongin 449701, South Korea
{monowar, rupam, spathan}@networking.khu.ac.kr, cshong@khu.ac.kr

## ABSTRACT

Heterogeneous applications could be assimilated within the same wireless sensor network with the aid of modern motes that have multiple sensor boards on a single radio board. Different types of data generated from such types of motes might have different transmission characteristics in terms of priority, transmission rate, required bandwidth, tolerable packet loss, delay demands etc. Considering a sensor network consisting of such multi-purpose nodes, in this paper we propose Prioritized Heterogeneous Traffic-oriented Congestion Control Protocol (PHTCCP) which ensures efficient rate control for prioritized heterogeneous traffic. Our protocol uses intra-queue and inter-queue priorities for ensuring feasible transmission rates of heterogeneous data. It also guarantees efficient link utilization by using dynamic transmission rate adjustment. Detailed analysis and simulation results are presented along with the description of our protocol to demonstrate its effectiveness in handling prioritized heterogeneous traffic in wireless sensor networks.

## Categories and Subject Descriptors
C.2.2 [Computer-Communication Networks]: Network Protocols

## General Terms
Algorithm, Design, Performance.

## Keywords
Heterogeneous, Congestion, Inter-Queue, Intra-Queue, Scheduler

## 1. INTRODUCTION
The sophistication of various communication protocols [1] and rapid advancements of Micro-Electro-Mechanical Systems (MEMS) technologies [2] have created a great opportunity for wide-spread utilizations of various innovative sensor network applications in near future. Today's sensor nodes are capable of sensing more than one parameter with the aid of multiple sensor boards mounted on a single radio board. Crossbow MICA2 mote [3] is an example of such type of sensor node. ExScal mote, an extension of MICA2 mote designed by CrossBow Inc. and Ohio State University also supports multiple sensing units [12], [13]. Instead of using multiple nodes with various functionalities [4], deploying such types of nodes might offer cost effective solutions for many applications. For example, a volcano monitoring application might require temperature, seismic, and acoustic data from the target location. Several applications could even run at the same time based on various types of data sent by the multi-purpose nodes in action. Different types of data in such a network might have different levels of importance and accordingly their transmission characteristics might differ.

In this paper we consider a wireless sensor network where all the deployed nodes are multi-purpose nodes and they generate heterogeneous traffic destined to the base station. Various types of data generated by the sensors have various priorities. Hence, it is necessary to ensure desired transmission rate for each type of data based on the given priority to meet the demands of the base station. In such a network, the sensor nodes could in fact generate simple periodic events to unpredictable bursts of messages. Congestion becomes even more likely when concurrent data transmissions over different radio links interact with each other or when the reporting rate to the base station increases. With the increase of number of nodes in the network, congestion might occur frequently. Therefore, it is necessary to have efficient mechanisms so that congestion could be controlled by ensuring balanced transmission rates for different types of data. Bearing these requirements in mind, here we propose an efficient congestion control protocol for a multi-purpose sensor network, so that the desired transmission rates of prioritized data could be maintained throughout the network's lifetime. In addition, our scheme also targets to ensure efficient link utilization by using a node-priority based hop-by-hop dynamic rate adjustment technique.

The rest of the paper is organized as follows; Section 2 states our motivation for this work and the relevant works, Section 3 presents our network model, design goals, and preliminaries for PHTCCP, Section 4 presents the description of PHTCCP. The MAC layer features are presented in section 5. Detailed analysis and simulation results are presented in Section 6 to show the efficiency of PHTCCP, and finally Section 7 concludes the paper mentioning the achievements of this work with future research directions.

## 2. MOTIVATION AND RELATED WORKS
A number of previous works have addressed the issue of congestion control in wireless sensor networks [5]. But most of the works have dealt with the rate control for homogeneous applications. In fact, no other work except STCP [6] has considered the use of multi-purpose sensor nodes in the network. Sensor Transmission Control Protocol (STCP) is a generic, scalable and reliable transport layer protocol where a majority of the functionalities are implemented at the base station. However, STCP has some drawbacks like; i) it doesn't provide any explicit and detailed mechanism for controlling congestion in the network, ii) the ACK/NACK based reliability mechanism might not be feasible for wireless sensor networks in terms of delay and memory usage.

CODA (Congestion Detection and Avoidance) [7] is a congestion mitigation strategy which uses both buffer occupancy and channel load for measuring congestion levels in the network. It uses two strategies for handling both persistent and transient congestions. CODA performs rate adjustment through traditional TCP-like AIMD (Additive Increase Multiplicative Decrease) mechanism and thus often leads to the occurrence of packet loss.

Fusion [8] detects congestion by measuring the queue length. It controls congestion by combining three techniques; hop-by-hop flow control, source rate limiting, and prioritized MAC. Fusion claims to achieve good throughput and fairness at high offered load.

In [9], the authors proposed a congestion control technique which uses packet service time to infer the available service rate and therefore detects congestion in each intermediate sensor node. It controls congestion in a hop-by-hop manner and uses exact rate adjustment based on its available service rate and number of child nodes. However, it cannot utilize the available link capacity efficiently when some nodes are in sleep state.

Siphon [10] is another congestion mitigation scheme which detects congestion using queue length. But instead of using any rate adjustment technique, it uses traffic redirection to mitigate congestion. Very recently a node-priority based congestion control mechanism, PCCP [11] has been proposed for WSN. It introduces an efficient congestion detection technique addressing both node and link level congestion. However, it doesn't have any mechanism for handling prioritized heterogeneous traffic in the network.

The scarcity of an efficient congestion control protocol for handling diverse data within a single node motivate us to design and develop PHTCCP. Our protocol uses an effective congestion detection mechanism which addresses both node and link level congestions. It also ensures application priority based rate control as well as high utilization of link capacity.

## 3. DESIGN GOALS, NETWORK MODEL, AND PRELIMINARIES

This section states the design considerations, network model, and preliminaries for PHTCCP which we have taken into account while constructing our algorithm. This is to be mentioned that, throughout the paper the terms *rate control* and *congestion control* are used interchangeably.

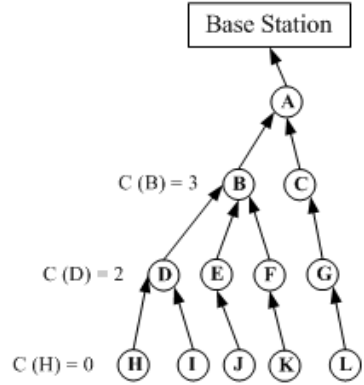

**Figure 1. Network model.**

### 3.1 Network Model and Assumptions

We consider a wireless sensor network where hundreds and thousands of multi-purpose nodes are deployed over a specific target area. All the nodes are equipped with the same number of different sensor boards (for our case we consider three types) mounted on a single radio board. Each of the nodes can sense different types of data at the same time and sends those to the base station. Figure 1 shows a model for our network where the data are forwarded in a many-to-one multihop single path routing fashion. We consider congestion control in such a network setting.

All nodes are supposed to use CSMA (Carrier Sense Multiple Access) like MAC protocol. The details of our MAC protocol features are presented in section 5. We assume that the network structure and the routes to the base station have been established by using some efficient routing protocol. While establishing the structure of the network, the base station dynamically assigns individual priority for each type of data. During forwarding heterogeneous data towards the base station, each sensor node transmits route data of its children nodes as well as its own generated data. So, at any given time a sensor node may act both as a source node and a forwarding node. When a sensor node transmits its data to the upstream direction, then it is called a child node and its immediate upstream node is called its parent. Each link between any parent and child is bidirectional that is if the child gets its parent within its transmission range, the parent also gets the child within its transmission range. We denote the number of the child nodes for a parent node $K$ as $C(K)$. As shown in Figure 1, node B has 3 children, D has 2 and node H doesn't have any child node. For each node in the network, there is a single path to reach to the base station.

### 3.2 Queuing Model

Figure 2 depicts the queuing model in a particular node. We assume that each node *i* has *n* number of equal sized priority queues for *n* types of sensed data. For example, a sensor node might sense temperature, light, and humidity at the same time. In such a case, there are 3 separate queues for each type of data. The number of queues in a node depends on the application requirements. As shown in Figure 2, a classifier has been provisioned in the network layer. The purpose of putting this classifier is to classify heterogeneous traffic either generated by the same node or incoming from other nodes. Based on the type of data, they are placed in the apposite queue.

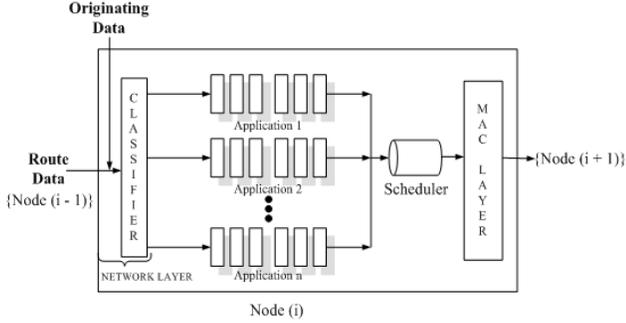

**Figure 2. Queuing model for a particular node.**

## 3.3 Definitions
Now we note down some important definitions that will be used throughout the rest of the paper.

*Originating Rate*: The rate at which a sensor node originates data. $R_{or}^i$ denotes the originating rate for node *i*.

*Scheduling Rate*: As shown in Figure 2, in our model, a scheduler is placed between the network layer and MAC layer. The scheduling rate is defined as how many packets the scheduler schedules per unit time from the queues. It is denoted by $R_{sch}^i$. The scheduler forwards the packets to the MAC layer from which the packets are delivered to the next node (i.e., *i*+1) along the path towards the base station.

*Average Packet Service Rate*: This is the average rate at which packets are forwarded from the MAC layer. It is denoted by $R_s^i$.

*Inter-Queue Priority*: We mentioned earlier that the base station assigns the priorities for heterogeneous traffic. Therefore, each data queue shown in Figure 2 has its own priority. This is termed as Inter-Queue priority. The scheduler schedules the queues according to the Inter-Queue priority. It decides the service order of the data packets from the queues and manages the queues according to their priorities. This ensures the data with higher priority to get higher service rate.

*Intra Queue Priority*: All the queues shown in Figure 2 are priority queues. Priority queues are used for giving the route data more priority than originating data. The reason behind this is that, as route data have already traversed some hop(s), their loss would cause more wastage of network resources than that of the originating (i.e., source) data. Hence, it would be better to forward those as soon as possible after receiving from the immediate downstream node. We term this type of priority as Intra-Queue priority. The classifier can assign the priority between the route data and originating data by examining the source address in the packet header.

## 4. PRIORITIZED HETEROGENEOUS TRAFFIC-ORIENTED CONGESTION CONTROL PROTOCOL (PHTCCP)
In this section, we describe our proposed protocol. The major goals for our scheme are: i) Generating and transmitting the heterogeneous data on priority basis. ii) Adjusting the rate while congestion occurs and ii) to ensure efficient link capacity utilization when some nodes in a particular route are inactive or in sleep mode. PHTCCP uses Weighted Fair Queuing (WFQ) for scheduling. Here, we illustrate PHTCCP in detail using several subsections to address the issues of congestion detection, notification, and mitigation.

### 4.1 Congestion Detection Method
In our protocol, we use packet service ratio $r(i)$ to measure the congestion level at each node *i*. Packet service ratio is defined as the ratio of average packet service rate ($R_s^i$) and packet scheduling rate ($R_{sch}^i$) in each sensor node *i* that is,

$$r(i) = R_s^i / R_{sch}^i \quad (1)$$

Here, the packet service rate $R_s^i$ is the inverse of packet service time $t_s^i$. The packet service time $t_s^i$ is the time interval when a packet arrives at the MAC layer and when it is successfully transmitted towards the next hop. $t_s^i$ includes packet waiting time, collision resolution, and packet transmission time at MAC layer. In equation 1, in order to obtain $R_s^i$, the average packet service time, $t_s^i$ is calculated using exponential weighted moving average formula (EWMA). By using EWMA, $t_s^i$ is updated each time a packet is forwarded as,

$$t_s^i = (1 - w_s) \times t_s^i + w_s \times inst(t_s^i) \quad (2)$$

where, $inst(t_s^i)$ is the instantaneous service time of the packet that has just been transmitted and $w_s$ is a constant where, $0 < w_s < 1$.

The packet service ratio reflects the congestion level at each sensor node. When this ratio is equal to 1, the scheduling rate is equal to the forwarding rate (i.e., average packet service rate). When this ratio is greater than 1, the scheduling rate is less than the average packet service rate. Both of these cases indicate the decrease of the level of congestion. When it is less than 1, it causes the queuing up of packets at the queue. This also indicates link level collisions. Thus, the packet service ratio is an effective measure to detect both node level and link level congestion.

### 4.2 Implicit Congestion Notification
PHTCCP uses implicit congestion notification. Each node *i* piggybacks its packet scheduling rate $R_{sch}^i$; total number of child nodes, $C(i)$; number of active child nodes at time *t*, $A_t(C(i))$; and the weighted average queue length of its active child nodes in its packet header. Because of the broadcast nature of wireless channel, all the child nodes of node *i* overhear the congestion notification information. Whenever the value of $r(i)$ goes below a certain threshold (depends on the application requirement), rate adjustment procedure is triggered.

## 4.3 Rate Adjustment

PHTCCP uses hop-by-hop rate adjustment which ensures that heterogeneous data reach to the base station at their desired rates. The output rate of a node is controlled by adjusting the scheduling rate, $R_{sch}^i$. We have stated earlier that the information of packet service ratio for congestion detection is piggybacked in the packet header along with other parameters. Each node *i* updates its scheduling rate if this ratio goes below the threshold or if there is any change in the scheduling rate of its parent node. The initial scheduling rate is set to $r_{sch}^{init}$.

**Table 1. Basic Notations used in the paper**

| Notation | Description |
| --- | --- |
| $R_{sch}^i$ | Scheduling rate of node *i* |
| $R_{or}^i$ | Originating rate of node *i* |
| $R_{sch}^{p_i}$ | Scheduling rate of the parent of node *i* |
| $A_t(C(p_i))$ | Total number of active child nodes of the parent of node *i* at time *t* |
| $C(p_i)$ | Total number of child nodes of the parent of node *i* |
| $E(t)$ | Excess link capacity at time *t* |
| $\varphi_i(t)$ | Weight factor for node *i* at time *t* |
| $\alpha_j^i$ | Priority for the *j*th queue of node i, *where, j=1,2,..,n* |
| $q_j^i$ | Current queue length for *j*th queue of node *i, where, j=1,2,..,n* |
| $avg_i^q(t)$ | Weighted average queue length of node *i* at time *t* |
| $N$ | Number of queues in node *i* |

Before presenting the rate adjustment algorithm, we present the notations and their illustrations in Table 1. The entire rate adjustment algorithm is shown in Figure 4. The algorithm works as follows:

Each node *i* measures its scheduling rate by calling the *Calculate_Scheduling_Rate()* method. In this method, at first each node *i* calculates its packet service ratio. When this ratio is equal to 1, it means that the incoming rate of packets to the MAC layer is equal to the average packet service rate (that is the rate at which packets are forwarded from the MAC layer). This is the ideal case so that no congestion (node level or link level) occurs. In this case, the scheduling rate of *i*, $R_{sch}^i$ remains unchanged. $R_{sch}^i$ remains unchanged as long as the packet service ratio doesn't go below the specified threshold. In fact, when the packet service ratio ($r(i)$) is less than the specified threshold value (say noted by $\mu$), it indicates that the scheduling rate of packets is larger than the average packet service rate. In such a case, packets would be queuing up at the MAC layer buffer and might cause buffer overflow indicating congestion. To control congestion, in this case, the scheduling rate is reset (i.e., decreased) to the value of packet service rate.

When $r(i)$ reaches above 1, it indicates that the packet service rate is greater than the scheduling rate. In this case, the scheduling rate is increased using the equation, $R_{sch}^i = \beta * R_s^i$. Here the value of $\beta$ is chosen to a value smaller than but close to 1. In our protocol, it is set to 0.75.

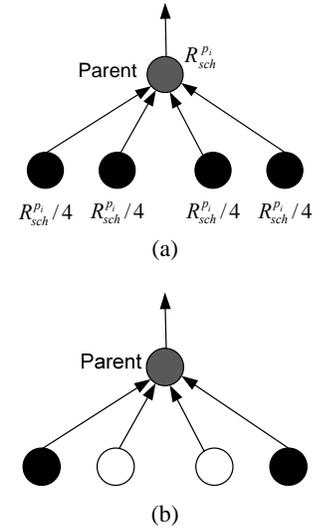

**Figure 3.** Any of the child nodes is termed as i, and the grey colored node is the parent of i, (a) All child nodes are active (black colored nodes) (b) Two child nodes are idle (white).

After determining the desired scheduling rate, each node *i* adjusts its own scheduling rate according to the scheduling rate of its parent node. This is done dynamically by calling the method *Dyn_Rate_Adj()*.

The rate adjustment depends on two cases:

When node *i* determines that all the child nodes of its parent (including itself) are active at time *t*, (in Figure 3(a), $A_t(C(p_i)) = C(p_i)$), then node *i* makes adjustment in its scheduling rate. In this case, each node *i* sets its scheduling rate equal to $1/C(p_i)$ th of its parent's scheduling rate. In Figure 3(a), if the scheduling rate of the parent node is $R_{sch}^{p_i}$, each child node has the scheduling rate, $R_{sch}^{p_i}/4$. This ensures that the total scheduling rate of all the child nodes is not greater than the scheduling rate of their parent node.

```
------------------------------------------------------------------------
Algorithm: Rate Adjustment
Input: Each node i;
Output: Scheduling rate R_sch^i, Originating rate R_or^i
```

**Initialization()**

$R_{sch}^i = r_{sch}^{init}; \quad r(i) = 1;$

**Calculate_Scheduling_Rate(** $R_{sch}^i, R_s^i$ **)**

$r(i) = R_s^i / R_{sch}^i$

If $r(i) < \mu$ then $R_{sch}^i = R_s^i$  End If

If $r(i) > 1$ then $R_{sch}^i = \beta * R_s^i$  End If

return $R_{sch}^i$

**Dyn_Rate_Adj(** $R_{sch}^{p_i}, A_t(C(p_i)), C(p_i), E(t)$ **)**

If $A_t(C(p_i)) = C(p_i)$ then $R_{sch}^i = R_{sch}^{p_i} / C(p_i)$
  End If

If $A_t(C(p_i)) < C(p_i)$ then $R_{sch}^i = R_{sch}^i + \varphi_i(t)E(t)$
  End If

**Calc_ExcessLinkCapacity(** $R_{sch}^{p_i}, A_t(C(p_i)), C(p_i)$ **)**

$$E(t) = \sum_{n=1}^{C(p_i)} R_{sch}^{p_i}/C(p_i) - \sum_{n=1}^{A_t(C(p_i))} R_{sch}^{p_i}/C(p_i)$$

return $E(t)$

**Calc_NodeWeightFactor(** $\alpha_j^i, q_j^i, A_t(C(p_i))$ **)**

$$avg_i^q(t) = \frac{\sum_{j=1}^{N} \alpha_j^i \times q_j^i}{N}$$

$$\varphi_i(t) = \begin{cases} \dfrac{avg_i^q(t)}{\sum_{i \in A_t(C(p_i))} avg_i^q(t)} & i \in A_t(C(p_i)) \\ 0 & otherwise \end{cases}$$

return $\varphi_i(t)$

**Calculate_SourceRate(** $R_{sch}^i, \alpha_i$ **)**

$$R_{or}^i = \frac{R_{sch}^i(t) * \alpha_i}{\alpha_1 + \alpha_2 + \cdots + \alpha_n}$$

return $R_{or}^i$

-------------------------------------------

**Figure 4. Rate Adjustment Algorithm.**

When node $i$ determines that some of the child nodes of its parent (i.e., its siblings) are idle that is when $A_t(C(p_i)) < C(p_i)$, it again adjusts its scheduling rate.

To achieve higher link utilization by taking advantage of excess link capacity, $E(t)$ is distributed to the active child nodes according to their weight factor $\varphi_i(t)$ at a particular time $t$. $\varphi_i(t)$ is determined dynamically using the *Calc_NodeWeightFactor()* method. Here the weight factor of the node depends on its weighted average queue length at time $t$.

The weighted average queue length is calculated by using the formula,

$$avg_i^q(t) = \frac{\sum_{j=1}^{N} \alpha_j^i \times q_j^i}{N} \quad (3)$$

Here $\alpha_j^i$ is the priority and $q_j^i$ is the length for queue *j* at time *t*. The weight $\varphi_i(t)$ reflects how the excess link capacity is to be allocated among the active nodes and is normalized such that,

$$\sum_{i \in A_t(C(p_i))} \varphi_i(t) = 1 \quad (4)$$

The excess link capacity is measured by using the procedure *Calc_ExcessLinkCapacity()*. It can be calculated just by subtracting the total scheduling rate of active child nodes from the total scheduling rate of all the child nodes.

After calculating the scheduling rate, each node $i$ updates their originating rate ($R_{or}^i$) according to the method *Calculate_SourceRate()*. The originating rate depends on the scheduling rate as well as on the priority for each type of data assigned by the base station.

## 5. TRAFFIC PRIORITY BASED MAC PROTOCOL

It is highly desirable in the sensor network environment that the MAC protocol operates without any centralized control. Thus, our MAC protocol is mainly based on distributed CSMA with RTS/CTS collision avoidance following the strategy of DCF (Distributed Coordination Function) mode of 802.11. The prioritization of traffic can be achieved by differentiating inter-frame-spacing (IFS) and back-off mechanisms. The basic idea is to assign short IFS and back-off to the higher priority traffic so that they can access the channel earlier than lower priority traffic [15] [16]. Hence, we adopt IEEE 802.11e [17] prioritization with some minor changes. The priority for each queue is mapped to one MAC priority class. Hence, each queue has different AIFS (Arbitration Inter Frame Space), CW (Contention Window), and PF (Persistence Factor) value according to its priority. This way, we can minimize the inter-node priority inversion such that higher priority packet in one node is not likely to be blocked by a lower priority packet in another node.

## 6. PERFORMANCE ANALYSIS AND SIMULATION

We have performed extensive simulation to evaluate the performance of PHTCCP in ns-2 [14]. We have determined a feasible threshold value of packet service ratio (μ) and found that the proposed protocol maintains a moderate queue length to avoid buffer overflow. We have also studied the application priority based throughput, bandwidth utilization, and energy efficiency of our proposed protocol.

The simulation settings for evaluating PHTCCP were as follows: 100 sensors were randomly deployed in $100 \times 100\ m^2$ sensor field. The transmission range of each sensor was set to 30 $m$. Maximum communication channel bit rate was 32 $kbps$. Each packet size was set to 33 bytes. The control packet size (RTS, CTS, and ACK) was set to 3 bytes. The weight used in the exponential weighted moving average calculation of packet service time (equation 2) was set to 0.1. We considered three sensing units (e.g., temperature, seismic, and acoustic) mounted on the single radio board for each node in which temperature flow is set the highest priority valued as 3; seismic reading is given the priority value 2, and the acoustic as 1. Each queue size was set to hold maximum 10 packets. That is, the total queue length for a node was 30 packets (10 packets for each queue). Through out the simulation, we used a fixed workload that consists of 10 sources and 1 sink. The initial originating rate was 4 pps (packets per second) and maximum originating rate was limited to 16 pps. We have compared our protocol with CCF[9] as it also performs the distributed rate adjustment of the child nodes based on the parent's transmission rate. Here, we have used the term buffer and queue interchangeably.

The IEEE 802.11e MAC protocol provided in ns-2 [17] simulator was used in our simulation. As a routing protocol Directed Diffusion [18] was used. The simulation was performed for 60 seconds.

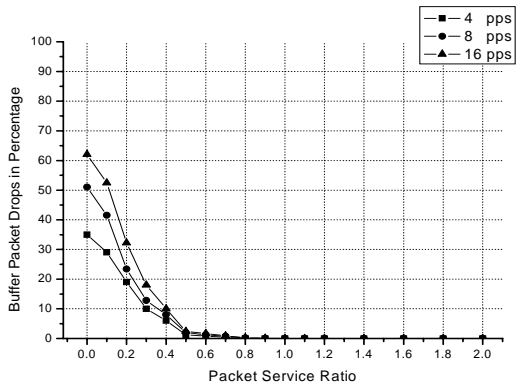

**Figure 5. Percentage of packet drops versus packet service ratio for different originating rates to determine the threshold value of μ.**

Figure 5 demonstrates how to determine the threshold of packet service ratio. It shows the percentage of buffer packet drops (irrespective of the traffic type) for different packet service ratios considering different packet originating rates. It's noticeable from the figure that the increase in the ratio reduces the percentages of packet drops. For different packet originating rates (*pps* – packet per second), the buffer packet drop percentage gradually goes below and reaches to an almost stable state (about 2%) when the packet service ratio becomes 0.5. This is a tolerable value before notifying any congestion. Hence, we set the value of the threshold ($\mu$) to 0.5.

Figure 6 illustrates the impact of packet service ratio over weighted average queue length, $avg_i^q(t)$ at the node closest to the sink. It shows that the weighted average queue length increases because of the increase of packet service ratio. This is because, increase in packet service ratio speeds up the packet service rate. In such case, scheduling rate should be increased in such a way that it doesn't cause any buffer overflow [As increasing the scheduling rate in the parent node also increases the scheduling rate of the child node through Dyn_Rate_Adj() method]. By setting the value of $\beta$ to 0.75, a moderate queue length could be maintained. We have run the simulation for 60 seconds and measured the weighted average queue length over time as shown in the Figure 7. This figure shows that the maximum weighted average queue length reaches to 9 packets and on an average it stays in between 3 to 5 packets throughout the simulation period. This indicates that PHTCCP maintains moderate queue length to avoid overflow.

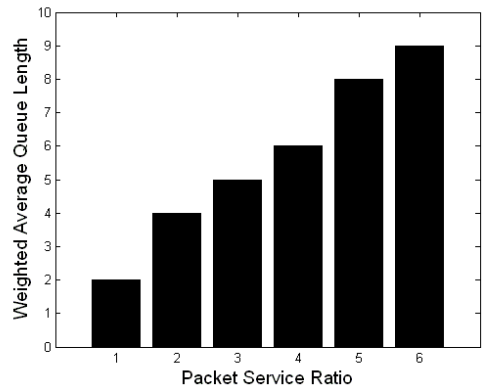

**Figure 6. Weighted Average queue length for different packet service ratio at the node near the sink.**

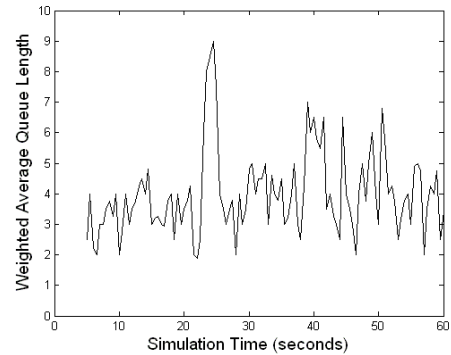

**Figure 7. Weighted Average queue length over time at the node near the sink reflecting the moderate queue length.**

We have analyzed the maximum memory requirements based on the queue sizes and number of sensing units considered onboard. The Figure 8 shows the maximum memory requirements for different packet sizes (considering 29 byte, 33 byte, 41 byte, and 64 byte packets). The memory requirements can be calculated by using the following equations:

$$M_r = N \times p_l \times q_l$$

where, $p_l$ is the packet length, $N$ is the total number of queues, and $q_l$ is the size of each queue. As we have considered three queues in total and each queue can contain maximum 10 packets, the memory requirements are 870, 990, 1230, and 1920 bytes for packet sizes 29, 33, 41, and 64 bytes respectively. Thus it shows that for packet size of 64 bytes, which is long enough for a sensor network application, the memory requirement is less than 2 KB. Hence, if a sensor mote has at least 4KB (4096 Bytes) onboard memory, the maximum memory occupancy would be less than 50% and on an average it is less than 30% which proves that our protocol could well be supported with current specifications of motes.

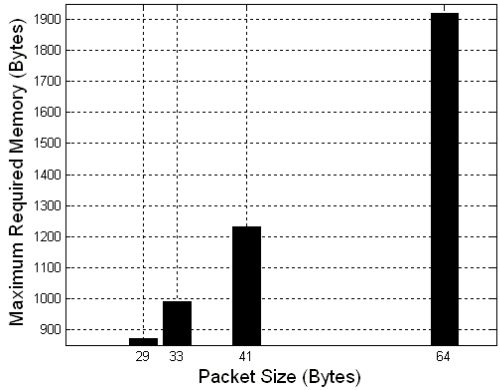

**Figure 8. Maximum memory requirements considering different packet sizes.**

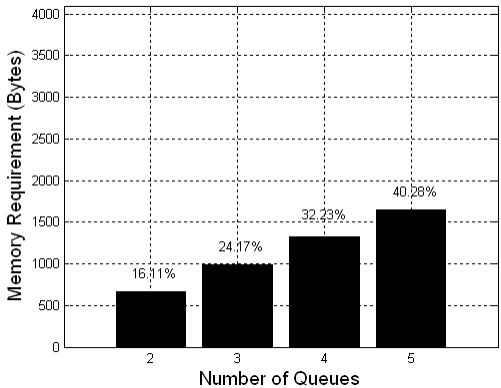

**Figure 9. Memory requirement (bytes) and percentage of memory allocation for different number of queues (Considering packet size of 33 bytes).**

Figure 9 shows the memory requirements for different number of queues considering 33 byte packets. It illustrates that with 33 bytes of packet size, even if we have simultaneously 5 different sensing units (i.e., 5 different queues), the protocol has 41% memory occupancy if the mote has at least 4 KB onboard memory. When the number of queues is 3, the occupancy is about 25% of total available onboard memory.

Figure 10 shows the number of different types of packets received by the base station over time. As per the priority given to the diverse data, the number of packets received i.e. the sink received highest number of temperature packets and then seismic packets and acoustic packets were the lowest in number throughout the simulation period.

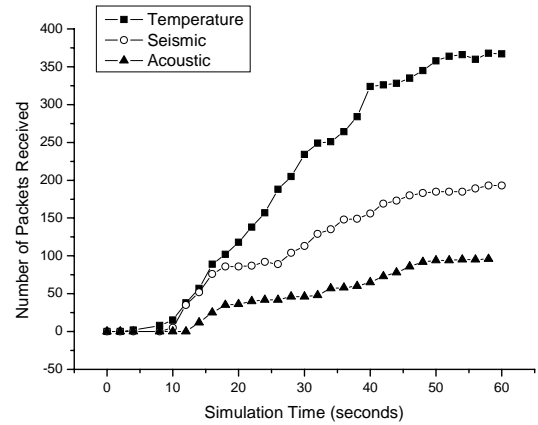

**Figure 10. Number of heterogeneous data received by the base station over time.**

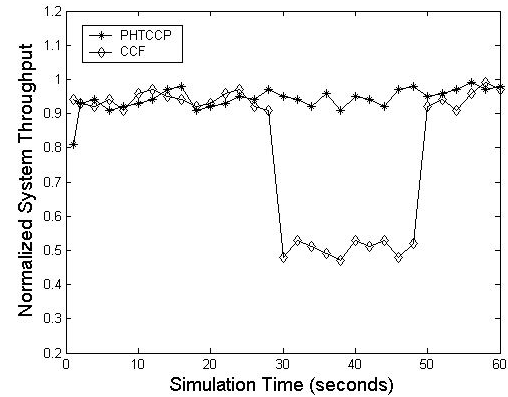

**Figure 11. Normalized system throughput over time.**

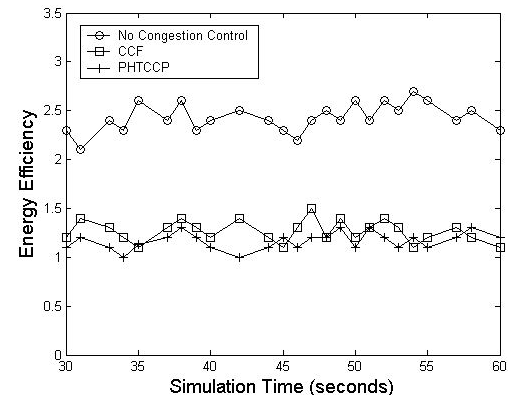

**Figure 12. Normalized system throughput over time.**

Figure 11 compares the normalized system throughput between PHTCCP and CCF [9]. The system bandwidth is normalized to 1. Within the time between 30 to 50 seconds, some nodes are set as idle. In that interval PHTCCP achieves higher system throughput than CCF since it allocates the excess link capacity to the active nodes.

In our simulation we define energy efficiency as:

$$\frac{T}{RH}$$

Where $T$ is the number of bytes transmitted in the whole network during a period of time, $R$ is the number of data bytes received by the base station during the same time and $H$ is the average number of hops a delivered packet travels. *A smaller value indicates better efficiency.* This measurement includes the actual transmission of data, the energy waste due to collision, and the energy waste due to packet drops.

Figure 12 compares the energy efficiencies in three cases: No Congestion Control, CCF, and PHTCCP. The figure shows that CCF and PHTCCP achieve much better energy efficiency than no congestion control because of lower packet drop rate. Again PHTCCP performs better than CCF.

## 7. CONCLUSION AND FUTURE WORKS

In this paper, we have presented PHTCCP, an efficient congestion control mechanism for heterogeneous data originated from multipurpose sensor nodes. We have demonstrated through simulation results and analysis that PHTCCP achieves, i) desired throughput for diverse data according to the priority specified by the base station, ii) high link utilization, iii) moderate queue length to reduce packet loss, iv) relatively low packet drop rate. Therefore PHTCCP is energy efficient and provides lower delay. It is also feasible in terms of memory requirements considering the configurations of today's multi-purpose motes. As our future work, we would like to work on integrating end-to-end reliability mechanism and improving fairness for PHTCCP.